\documentclass[twocolumn,showpacs,floatfix]{revtex4}
\usepackage{graphics}

\begin{document}

\title{Antisymmetric entangled two-photon states generated
in nonlinear GaN/AlN photonic-band-gap structures}

\author{Jan Pe\v{r}ina, Jr.}
\affiliation{Joint Laboratory of Optics of Palack\'{y} University
and Institute of Physics of Academy of Sciences of the Czech
Republic, 17. listopadu 50A, 772 07 Olomouc, Czech Republic}
\email{perinaj@prfnw.upol.cz}
\author{Marco Centini}
\author{Concita Sibilia}
\author{Mario Bertolotti}
\affiliation{Dipartimento di Energetica, Universit\`{a} La
Sapienza di Roma, Via A. Scarpa 16, 00161 Roma, Italy}
\author{Michael Scalora}
\affiliation{Charles M. Bowden Research Center, RD\&EC, Redstone
Arsenal, Bldg 7804, Alabama 35898-5000}

\begin{abstract}
The properties of an entangled two-photon state antisymmetric in
frequencies are studied. At a beam-splitter, two entangled photons
are perfectly anti-correlated. In addition, they cannot be
detected at the same time instant despite the fact that their
detection times are confined to a narrow time window, i.e. they
are temporally anti-bunched. Using nonlinear photonic-band-gap
structures made of GaN/AlN, two schemes for generating such states
are described.
\end{abstract}



\maketitle

\section{Introduction}

In the process of spontaneous parametric down-conversion, photons
are generated in pairs into a signal and an idler fields. Because
of uncertainty in the generated frequency and/or polarization
and/or wave-vector of the down-converted fields and because of
energy and momentum conservation, a state describing a photon pair
is entangled. This means that two photons constituting a photon
pair cannot be described separately and leads to an unusual
behavior of photon pairs.

Entanglement in frequencies manifests itself in the fact that both
photons appear in time very close to each other. The width of time
window characterizing the appearance of two entangled photons is
called an entanglement time and can be measured in a
Hong-Ou-Mandel interferometer \cite{Hong1987,Mandel1995}. A
Hong-Ou-Mandel interferometer is based on the fact that if two
photons with the same time profile (they need not be entangled,
see, e.g. \cite{Rarity1997,Santori2002} for interference of two
independent photons) reach two ports of a beam-splitter at the
same time instant, they both exit the beam-splitter from the same
output port. This occurs because the paths leaving one photon in
one output port and the other photon in the other output port
cancel due to destructive interference. We note that in fact it is
not the overlap of two photons at a beam-splitter, but their
overlap in the area of detectors measuring a coincidence count
that is required to observe this behavior \cite{Pittman1996}.
However, in the most of experimental setups, an overlap at a
beam-splitter insures also overlap at the detectors. This
correlation (coalescence) is unusual in classical physics in which
two statistically independent photons choose randomly their output
ports and so they may be observed in different output ports with a
probability of 50 percent. In a Hong-Ou-Mandel interferometer, a
mutual time delay is introduced between two photons that decreases
the mutual overlap of the photons. Subsequently, an entanglement
time can be determined from the width of a typical dip in the
fourth-order coincidence-count interference rate. In order to
observe 100 percent visibility of the interference pattern, two
photons have to be identical (or perfectly indistinguishable). For
this reason, a lot of attention has been paid to develop methods
for the generation of photon pairs containing identical signal and
idler photons \cite{Branning1999,Atature2000}. Considering usual
entangled two-photon states (i.e. states more-less symmetric in
their independent variables), it may be deduced from the shape of
the fourth-order coincidence-count interference pattern that the
probability of detecting an idler photon at a given time instant
is constant (in cw regime) in a certain time window provided the
signal photon has been detected at a given time instant,  or
decreases (for femtosecond pumping) as the difference of two time
instants increases.

The above mentioned picture changes when two photons are generated
in a state antisymmetric in a variable describing this state.
Provided that two photons are generated in an antisymmetric
polarization Bell state $ | \psi^- \rangle $ and enter a
beam-splitter in different input ports they exit the beam-splitter
in different output ports (see, e.g. in \cite{Branning1999}), and
we speak about anti-correlation (anti-coalescence) of two photons.
This property is exploited in many teleportation schemes
\cite{Braunstein1995}.

If two photons are generated in a state antisymmetric in
frequencies, they are anti-correlated at a beam-splitter. They
also show temporal anti-bunching, i.e. if a signal photon has been
detected at a given time instant, then the probability of
detecting an idler photon at a given time instant increases with
the increasing difference of two time instants for small values of
the difference (see Fig.~\ref{fig3} later). These states are
investigated in detail in this paper.

States antisymmetric in frequencies represent a temporal
(spectral) analogy of the states that show spatial anti-bunching
as a consequence of antisymmetry of their two-photon amplitude
with respect to the exchange of the signal- and idler-field wave
vectors. Several methods for the generation of spatially
anti-bunched entangled states have been proposed and
experimentally verified
\cite{Nogueira2001,Nogueira2002,Caetano2003,Nogueira2004}.

We note that entangled two photon states generated in spontaneous
parametric down-conversion have been essential for numerous
important experiments; we mention testing of Bell and other
nonclassical inequalities \cite{Perina1994,Bovino2004,Bovino2005},
quantum teleportation \cite{Bouwmeester1997}, quantum cloning
\cite{DeMartini2000}, and generation of
Greenberger-Horne-Zeilinger states \cite{Bouwmeester1999} to name
a few. Also, important applications of entangled two-photon states
have been developed \cite{Lutkenhaus2000,Migdal1999}.

\section{Entangled two-photon states antisymmetric in frequencies}

In an entangled two-photon state antisymmetric in frequencies
signal- and idler-field spectra may be considered as composed of
two peaks of equal heights, and ideally no field is generated at
the central frequencies of the signal and idler fields. In other
words, their two-photon amplitude is antisymmetric with respect to
the signal-field frequency and idler-field frequency exchange. We
note that methods for controlling correlations between the signal-
and idler-field frequencies have been suggested
\cite{Booth2002,Torres2005}.

Nonlinear photonic-band-gap structures
\cite{Joannopoulos1995,Sakoda2005,Bertolotti2001} represent a
suitable source of entangled two-photon states antisymmetric in
frequencies. The wave-function $ |\psi\rangle^{(2)}_{s,i} $ of a
state generated in a photonic-band-gap structure can be naturally
decomposed into four contributions that take into account the
propagation directions of the generated photons
(\cite{Centini2005}, and details may be found in
\cite{PerinaJr2005}):
\begin{equation}    
 |\psi(t)\rangle_{s,i}^{(2)} = |\psi_{s,i}^{FF}(t) \rangle +
 |\psi_{s,i}^{FB}(t) \rangle + |\psi_{s,i}^{BF}(t) \rangle +
 |\psi_{s,i}^{BB}(t) \rangle,
\end{equation}
where the superscripts $ F $ and $ B $ distinguish forward and
backward propagation of down-converted photons with respect to an
incident pump beam. Introducing the probability amplitudes $
\phi^{mn}(\omega_s,\omega_i) $ of generating a signal photon at
frequency $ \omega_s $ and an idler photon at frequency $ \omega_i
$, the contributions $ |\psi_{s,i}^{mn}\rangle $ can be written as
follows:
\begin{eqnarray}   
 |\psi_{s,i}^{mn}(t)\rangle &=&
  \int_{0}^{\infty} \, d\omega_s  \int_{0}^{\infty} \, d\omega_i \;
  \phi^{mn}(\omega_s,\omega_i)
  \nonumber \\
 & & \mbox{} \hspace{-1.5cm}\times
  \hat{a}_{s_m}^{\dagger}(\omega_s)
  \hat{a}_{i_n}^{\dagger}(\omega_i) \exp(i\omega_s t) \exp(i \omega_i t)
  |{\rm vac} \rangle , \nonumber \\
 & & \hspace{1cm} m,n = F,B .
\label{2}
\end{eqnarray}
The symbols $ \hat{a}_{s_m}^\dagger $ and $ \hat{a}_{i_n}^\dagger $
stand for creation operators of a photon into a given mode of the
quantized optical field with vacuum state $ |{\rm vac} \rangle $.

For simplicity, we consider cw pumping with central frequency $
\omega_p^0 $. The probability amplitude $ \phi^{mn} $ of our
entangled photon pair can be written in the general form that
follows (see also \cite{Branning1999}):
\begin{eqnarray}   
 \phi^{mn}(\omega_s,\omega_i) &=&
 \delta(\omega_p^0-\omega_s-\omega_i)  \nonumber \\
  & & \mbox{} \times \left[
 f(\omega_s-\omega_s^0) - f(-\omega_s+\omega_s^0) \right];
\label{3}
\end{eqnarray}
the complex function $ f $ is determined by the geometry. This form of $
\phi^{mn} $ ensures that the signal- and idler-field spectra are
composed of two symmetrically positioned peaks having the same
heights.

An entangled photon pair in the time domain is conveniently
described using a two-photon amplitude $ {\cal A}^{mn} $
\cite{Keller1997,PerinaJr1999,DiGiuseppe1997} defined as:
\begin{eqnarray}    
  {\cal A}^{mn}(\tau_s,\tau_i)
   &=& \langle {\rm vac} |
  \hat{E}^{(+)}_{s_m}(t_0+\tau_s) \nonumber \\
 & & \mbox{} \times \hat{E}^{(+)}_{i_n}(t_0+\tau_i)
  |\psi^{mn}_{s,i}(t_0)\rangle ;
\label{4}
\end{eqnarray}
the symbols $ \hat{E}^{(+)}_{s_m} $ and $ \hat{E}^{(+)}_{i_n} $ denote
positive-frequency components of electric-field amplitude operators for
signal and idler fields. Considering the state described by the
probability amplitude $ \phi^{mn} $ in Eq.~(\ref{3}), we arrive
at:
\begin{eqnarray}   
 {\cal A}^{mn}(\tau_s,\tau_i) &=& \frac{\hbar \sqrt{\omega_s^0 \omega_i^0}}{
 2\sqrt{2\pi}^3\epsilon_0 c {\cal B}}
 \exp(-i\omega_s^0\tau_s) \exp(-i\omega_i^0\tau_i) \nonumber \\
 & &  \mbox{} \times \left[ \tilde{f}(\tau_s-\tau_i)
 - \tilde{f}(\tau_i-\tau_s) \right];
\label{5}
\end{eqnarray}
$ \hbar $ is the reduced Planck constant, $ \epsilon_0 $ is the
permittivity of vacuum, $ c $ the speed of light, and $ {\cal B} $
gives the area of the transverse profiles of the interacting
beams. The symbol $ \tilde{f} $ denotes the Fourier transform of
the function $ f $ and $ \omega_s^0 $ ($ \omega_i^0 $) means a
central frequency of the signal (idler) field. Provided that $
\tau_s = \tau_i $ in Eq.~(\ref{5}), the two-photon amplitude $
{\cal A}^{mn} $ is zero. Because $ |{\cal
A}^{mn}(\tau_s,\tau_i)|^2 $ gives the probability of simultaneous
detection of the signal photon at time $ \tau_s $ and the idler
photon at time $ \tau_i $, the signal photon and its idler twin
cannot be simultaneously detected at the same time instant.

Anti-correlation of entangled photons at a beam-splitter may be
deduced from the behavior of the normalized coincidence-count rate
$ R_n^{mn} $ \cite{PerinaJr2005} in a Hong-Ou-Mandel
interferometer. We derive the following expression for the
normalized coincidence-count rate $ R_{\rm n}^{mn} $ assuming a
state with the probability amplitude $ \phi^{mn} $ given in
Eq.~(\ref{3}):
\begin{equation}    
 R_{\rm n}^{mn}(\tau_l) = 1 + \frac{\Re\left[ \exp[-i(\omega_s^0
  - \omega_i^0)\tau_l] g(\tau_l) \right] }{ g(0) },
 \label{6}
\end{equation}
where
\begin{equation}  
 g(\tau) = \int_{-\infty}^{\infty} \, d\omega |f(\omega) -
 f(-\omega)|^2 \exp(-2i\omega\tau)
\label{7}
\end{equation}
and symbol $ \Re $ means the real part of an expression. If the
mutual time delay $ \tau_l $ between two photons in a pair is
zero, i.e. both photons perfectly overlap at the beam-splitter, $
R_n = 2 $ according to Eq.~(\ref{6}). This means that both
entangled photons must leave the beam-splitter from different
output ports in order to have twice the coincidence counts
per second compared to the case of two independent photons ($
\tau_l \rightarrow \infty $).

\section{Two schemes for the generation of an entangled two-photon state
antisymmetric in frequencies in GaN/AlN structures}

Photonic-band-gap structures made of GaN/AlN (for characteristics,
see, e.g., \cite{Sanford2005}) offer two different schemes for
generating such states. The first scheme relies on the vectorial
character of spontaneous parametric down-conversion and uses
destructive interference between two paths that differ in
polarization properties. The second scheme is based on the
observation \cite{PerinaJr2005} that an efficient photon-pair
generation occurs at frequencies that lie at resonance peaks of
the linear transmission spectrum of the structure. This suggests a
design of a photonic-band-gap structure such that photon pairs are
efficiently generated at two neighboring transmission peaks.

\begin{figure}    
 \resizebox{0.5\hsize}{!}{\includegraphics{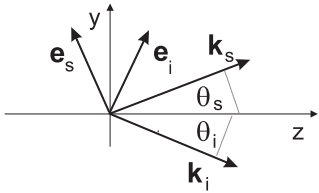}}
 \caption{Vectorial scheme of a generated photon pair. A signal (idler) photon
 with wave vector $ {\bf k}_s $ ($ {\bf k}_i $) propagates with p-polarization
 (in yz plane) along the angle
 $ \theta_s $ ($ \theta_i $) and is linearly polarized along vector $ {\bf e}_s $
 ($ {\bf e}_i $).}
\label{fig1}
\end{figure}

As an example, we consider a structure composed of 25 nonlinear
layers of GaN (thickness 110 nm) among which there are 24 linear
layers of AlN (thickness 60 nm). The boundaries are perpendicular to z
crystalographic axis of GaN that coincides with the z axis of the
coordinate system (see Fig.~\ref{fig1}). The structure is designed
for pumping at the wavelength of 395 nm at normal incidence.

\begin{figure}    
 \resizebox{0.8\hsize}{!}{\includegraphics{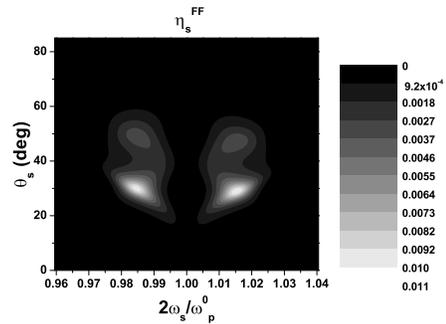}}
 \vspace{0mm}
 \caption{Photon-pair generation rate $ \eta_{s}^{FF} $ as a function
 of normalized signal-field frequency $ 2\omega_s/\omega_p^0 $ and
 angle $ \theta_s $ of signal-photon emission for forward-propagating
 signal and idler photons; cw pumping and p-polarization of all fields
 are assumed.}
\label{fig2}
\end{figure}

{\em The first scheme} based on destructive interference between
two quantum paths occurs in this structure when the pump, signal,
and idler fields are p-polarized (polarization vectors lie in yz
plane, see Fig.~\ref{fig1}) for photon pairs composed of
co-propagating photons (i.e., both photons are
forward-propagating, or backward-propagating). A two-peak
character of frequency dependence of the photon-pair generation
rate $ \eta_S^{FF} $ for forward-propagating photons is shown in
Fig.~\ref{fig2}.

Here, no photon pair with a signal photon at the central frequency
$ \omega_p^0/2 $ can be generated due to completely destructive
interference. GaN has two nonzero elements of nonlinear
susceptibility $ \chi^{(2)} $, $ \chi^{(2)}_{y,y,z} $ and $
\chi^{(2)}_{y,z,y} $, that are responsible for the generation of
photon pairs in this configuration. Considering, e.g., the element
$ \chi^{(2)}_{y,y,z} $, one photon is generated with polarization
along the y axis, while its twin has polarization along the z axis
(see Fig.~\ref{fig1}). The state of a photon pair with a signal
photon having wave vector $ {\bf k}_s $ and an idler photon with
wave vector $ {\bf k}_i $ is created by interference of two paths:
either a photon with polarization along the y axis becomes a
signal photon and then a photon with polarization along the z axis
has to be an idler photon, or vice versa. The probability
amplitudes of these two paths have a different sign due to the
vectorial character of nonlinear interaction (see Fig.~\ref{fig1};
the z components of the polarization vectors of the signal and
idler photons differ in sign) and as a result no photon pair can
be generated at the degenerate frequencies of the down-converted
fields (for $ \omega_s = \omega_i = \omega_p^0/2 $) due to
symmetry. Photon pairs are then generated symmetrically around the
degenerate central frequencies (see Fig.~\ref{fig2}).

\begin{figure}    
 \resizebox{0.5\hsize}{!}{\includegraphics{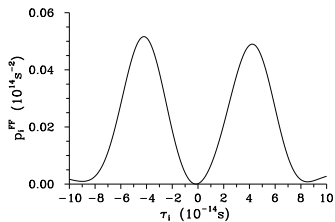}}
 \caption{Probability $ p_i^{FF} $ of detecting an idler photon at
 time $ \tau_i $ provided that its signal photon is detected
 at time $ \tau_s = 0 $~s is shown for forward-propagating down-converted
 photons with a signal photon emitted along the angle $ \theta_s = 30 $~deg,
 cw pumping and p-polarization of all fields are assumed.}
\label{fig3}
\end{figure}

The two photons in this entangled state cannot be detected at the
same time instant despite the fact that their possible detection
times are confined within a sharp time window (of typical duration
in hundreds of fs), as documented in Fig.~\ref{fig3}, where the
probability $ p_i^{FF}(\tau_i) $ of detecting an idler photon at
time $ \tau_i $ is plotted, provided that the signal photon is
detected at a given time $ \tau_s $.

\begin{figure}    
 \resizebox{0.5\hsize}{!}{\includegraphics{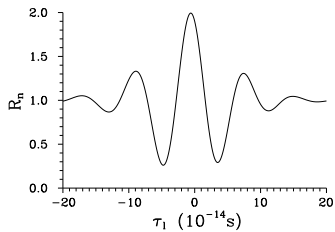}}
 \caption{Normalized coincidence-count rate $ R_{\rm n} $ as a function of
 relative time delay $ \tau_l $ in Hong-Ou-Mandel interferometer for
 forward-propagating signal
 and idler photons for the angle of signal-field
 emission $ \theta_s = 30 $~deg. All fields have p-polarization and
 cw pumping is assumed.}
\label{fig4}
\end{figure}

Perfect anti-correlation of two entangled photons in this state at
a beam-splitter is evident from the normalized coincidence-count
rate $ R_{\rm n} $ in a Hong-Ou-Mandel interferometer shown in
Fig.~\ref{fig4}, where for $ \tau_l = 0 $~s the coincidence-count
rate $ R_{\rm n} $ for both photons propagating forward is two
times greater in comparison with that for two uncorrelated photons
($ \tau_l \rightarrow \infty $). Oscillations in the
coincidence-count rate $ R_{\rm n} $ around the main peak in
Fig.~\ref{fig4} indicate the difference of the central frequencies
of two peaks.

Generation of these states occurs due to the fact that the
nonlinear GaN has a wurtzite structure. We note that materials
with a cubic symmetry cannot provide these states.

\begin{figure}    
 \resizebox{0.8\hsize}{!}{\includegraphics{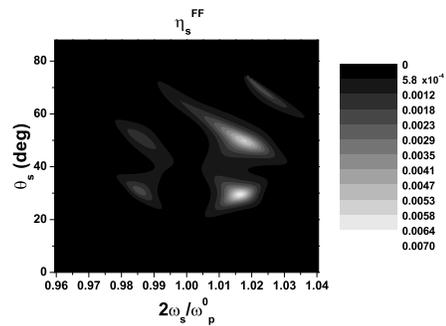}}
 \vspace{0mm}
 \caption{Photon-pair generation rate $ \eta_{s}^{FF} $ as a function
 of normalized signal-field frequency $ 2\omega_s/\omega_p^0 $ and
 angle $ \theta_s $ of signal-photon emission for forward-propagating
 signal and idler photons; cw pumping, p-polarized signal field, and
 s-polarized idler and pump fields are assumed.}
\label{fig5}
\end{figure}

The generation of an antisymmetric entangled photon pair along
{\em the second scheme} occurs in the considered structure in the
configuration with s-polarized idler and pump fields (their
polarization vectors are orthogonal to yz plane) and p-polarized
signal field. Pumping is at normal incidence. Photon pairs are now
generated owing to a nonzero value of the element $
\chi^{(2)}_{x,z,x} $. A two-peak structure of the photon-pair
generation rate $ \eta^{FF}_s $ with maxima at signal-field
emission angles $ \theta_s $ equal to 30 and 50 degrees (see
Fig.~\ref{fig5}) has its origin in the properties of the
photonic-band-gap structure that allow an efficient generation for
frequencies lying in the areas around the transmission peaks of
the linear transmission spectrum. However, different generation
rates for frequencies around different peaks exist. This is caused
by the fact that waves with frequencies around different
transmission peaks have a different degree of localization
(density of modes) inside the structure, and this affects the
efficiency of the nonlinear process. As a consequence, a
photonic-band-gap structure cannot provide the entangled state in
its perfect form. For instance, the number of coincidences for $
\tau_l = 0 $~s in a Hong-Ou-Mandel interferometer exceeds that for
$ \tau_l \rightarrow \infty $ by cca 60~\% for the considered
structure and $ \theta_s = 50 $ deg.

Considering s-polarized pump field at normal incidence and
detecting a signal photon as well as an idler photon in
polarization directions rotated by 45 degrees (with respect to
p-polarization direction) for $ \theta_s= 30 $ and $ 50 $~deg, we
have again the antisymmetric entangled state generated along the
first scheme. Now due to destructive interference of two
contributions originating in elements $ \chi^{(2)}_{x,z,x} $ and $
\chi^{(2)}_{x,x,z} $. This means that the effect of
anti-correlation at a beam-splitter and temporal anti-bunching is
perfect. We note that the considered structure with s-polarized
pumping at normal incidence represents a source of photon pairs
entangled in polarization, i.e. either a signal photon is
s-polarized and its twin p-polarized, or vice versa.

\begin{figure}    
 {\raisebox{3.3 cm}{a)} \hspace{5mm}
 \resizebox{0.8\hsize}{!}{\includegraphics{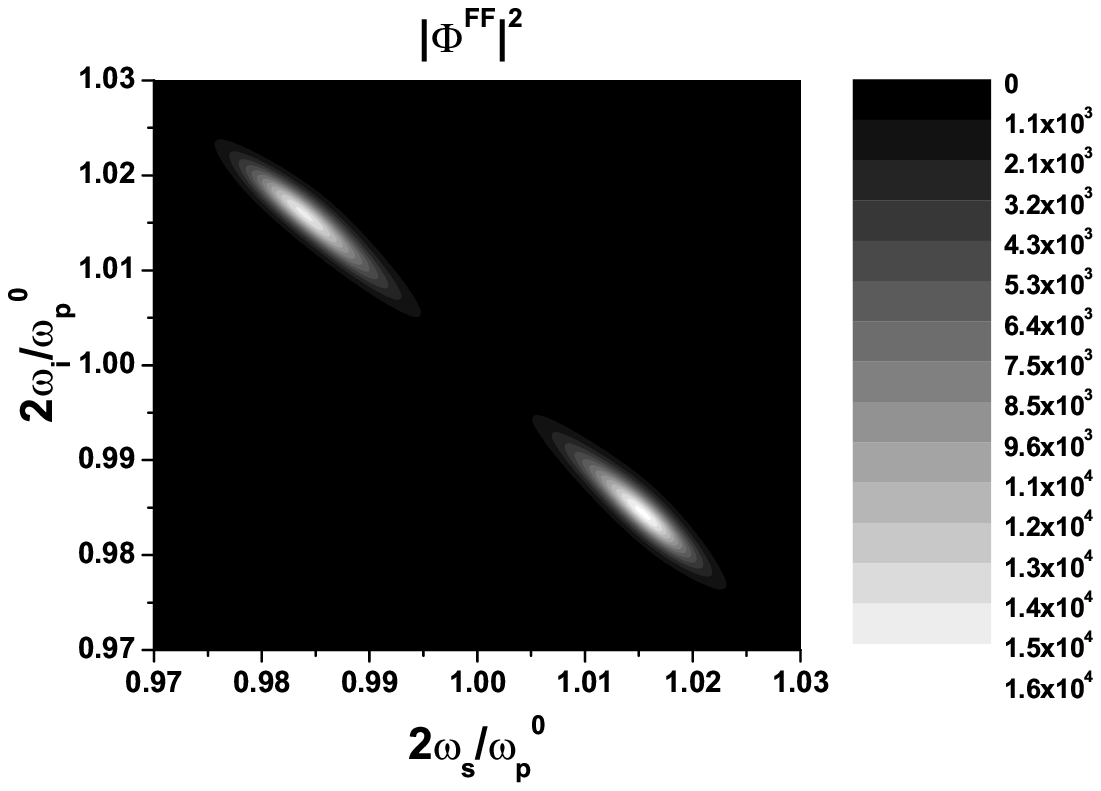}}

 \vspace{0mm}
 \raisebox{3.3 cm}{b)} \hspace{5mm}
 \resizebox{0.8\hsize}{!}{\includegraphics{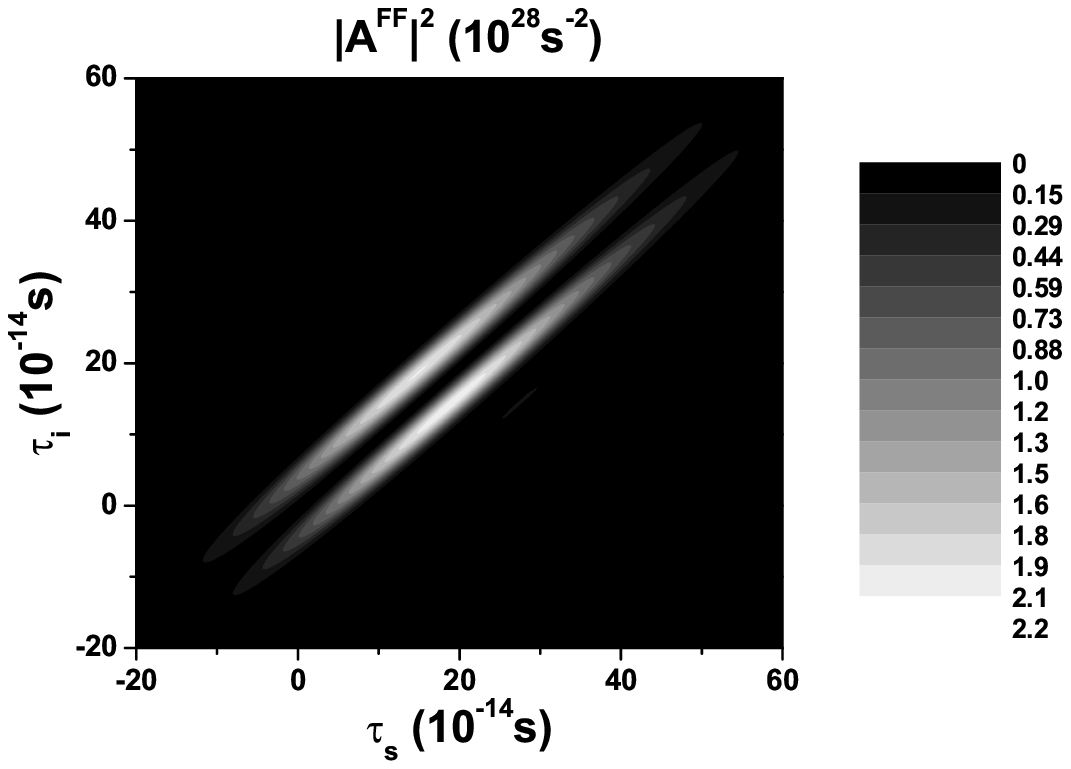}}}
 \vspace{0mm}

 \caption{Probability $ |\Phi^{FF}|^2 $ of generating a two-photon state with
 a signal photon at normalized frequency $ 2\omega_s/\omega_p^0 $ and
 an idler photon at normalized frequency $ 2\omega_i/\omega_p^0 $
 (a) and probability $ |{\cal A}^{FF}|^2 $ of detecting a signal photon at
 time $ \tau_s $ and an idler photon at time $ \tau_i $ (b) for forward-propagating
 down-converted photons and the angle of signal-field emission $ \theta_s = 30 $~deg.
 All fields are p-polarized and pumping by a
 gaussian pulse with 200-fs duration is assumed. Normalization of the
 probabilities is such that one photon pair is emitted within the plotted ranges.}
\label{fig6}
\end{figure}

{\em Pumping} the structure {\em with an ultrashort pulse} causes
the above described properties of the generated entangled photons
not only to survive, but also to be enhanced. For example, in
Fig.~\ref{fig6}a we depict the probability $ |\phi^{FF}|^2 $ of
generating a signal photon at the normalized frequency $
2\omega_s/\omega_p^0 $ together with its twin at the normalized
frequency $ 2\omega_i/\omega_p^0 $ for a gaussian pump pulse with
duration 200~fs. Splitting of the generated photon-pair (using the
first scheme) into two separated spectral regions is clearly
visible. On the other hand, the squared modulus of the two-photon
amplitude $ {\cal A}^{FF} $ in Fig.~\ref{fig6}b shows that
temporal anti-bunching of entangled photons is even more
pronounced in the pulsed regime. We note that the down-converted
fields now occur in the form of pulses with a typical duration in
the hundreds of fs.

\section{Conclusion}

In summary, we have shown that entangled two-photon states
antisymmetric with respect to the exchange of frequencies of the
signal and idler fields can be generated in nonlinear
photonic-band-gap structures. The photons comprising a pair
exhibit anti-correlation at a beam-splitter. Despite the fact that
both photons exist within a narrow time window, they cannot be
detected at the same time instant. Photonic-band-gap structures
offer two different ways for their generation: one exploits
vectorial character of the nonlinearly interacting fields together
with destructive interference between two quantum paths, the other
is based upon the generation of two adjacent transmission peaks.
These properties might be useful in future quantum-information
protocols \cite{Bouwmeester2000}. Especially splitting of an
entangled two-photon state into two perfectly separated spectral
regions even for femtosecond pumping is promising for further
considerations. Experimental realization of these states using
GaN/AlN structures is in progress.

\acknowledgments Support by projects COST OC P11.003,
MSM6198959213, 1M06002, and  AVOZ 10100522 of the Czech Ministry
of Education as well as support coming from cooperation agreement
between Palack\'{y} University and University La Sapienza in Rome
are acknowledged.

\end{document}